# Transport across meso-junctions of highly doped Si with different superconductors

Pradnya Parab, Sangita Bose*

*School of Physical Sciences, UM-DAE Center for Excellence in Basic Sciences, University of Mumbai, Kalina Campus, Mumbai 400098, India*

Email: **sangita@cbs.ac.in**

*Abstract:*

*We studied the transport properties of meso-junctions of semiconducting (Sm) highly doped Si with different superconductors (Sc) through point contact Andreev reflection (PCAR) spectroscopy. Spectra of low transparency point contacts between Si and In showed an enhancement in the superconducting energy gap of In. This was due to the effect of an additional gap arising from the Schottky barrier at the Sm-Sc interface. For higher transparency Si-Nb and Si-Pb point contacts, no gap enhancement was observed though there were weak sub gap features. These were due to proximity induced interface superconductivity known to occur for Sm-Sc junctions of high transparency.*

**Introduction**

Carrier transport in semiconductor (Sm)-superconductor (Sc) junctions has been probed extensively in the last decade[1,2,3,4,5,6,7,8,9,10,11]. In Sm-Sc junctions with high transparency, a finite super-current flow is detected which is understood in terms of Andreev reflection and the superconducting proximity effect (SPE)[3,4,5]. Furthermore, presence of Andreev scattering at the interface are shown to affect the current-voltage characteristics of the junctions. Recently there has been considerable interest in using the SPE to search for proximity induced superconductivity in topological insulators and semi-metals and also hunt for the Majorana modes[12,13]. It is proposed by Sau, Lutchyn, Tiwari and Das Sarma that such modes might exist at a contact of a Sc to a Sm nanowire[14,15]. Following this, some experimental signatures of the emergence of zero energy modes have been observed which has prompted researchers to hunt for more conclusive proof of the Majorana modes[16]. In recent times Sm-Sc hybrid devices have also been successfully fabricated which have shown promising applications as superconducting light emitting devices[17,18], waveguide amplifiers[19] or in fields of quantum information[20]. In most of these applications, the devices have exploited the phenomena of Andreev reflection.

Andreev reflection and SPE at the Sm-Sc interfaces is found to depend on the doping of the Sm and on the interface cleanliness and transparency[21,22,23]. Point contact spectroscopic technique has been employed to study Andreev reflection for Sm-Sc interfaces, where the Sm-Sc contact is primarily done by micro-fabrication techniques in thin film geometry. This has been used to study systems like Si/Nb[3,8,9], GaAs/Sn[24], InAs-AlSb/Nb[25], GaAs/Nb[10,26], InAs-Pb[27] *etc* where non-equilibrium pair currents have been detected in the Sm. In most of the earlier work, multiple peak structure was observed in the conductance spectra of the junctions. Interestingly, in some reports of diffusive Sm-Sc contacts, reflection less tunneling was observed which gave rise to huge enhancements of zero bias conductance (ZBC) which is considered a classic signature of phase coherence[28]. In some other reports, the voltage at which the coherence peaks in the conductance spectra was observed was much greater than the voltage corresponding to $\Delta_{SC}$ (energy gap of the superconductor) with no multiple peaks[10]. The point contact Andreev reflection (PCAR) spectra was theoretically modelled using the Blonder-Tinkham-Klapwijk (BTK) model which assumes a delta like potential barrier at the interface of the Sm-Sc. It provides an analytical approach for calculating the transport properties based on the Bogoliubov–de Gennes (BdG) equations[29] according to which it is expected to see the peaks in the conductance spectra at a voltage of ~ $\Delta_{SC}/e$. Lissitski *et. al.* further showed the effect of the Schottky barrier formed at the interface of the Sm-Sc interface.[30] They predicted that the peaks in the conductance spectra should appear at voltages greater than $\Delta_{SC}/e$. Through their simple model, they showed that the presence of the Schottky barrier reduced the charge screening resulting in a gap in the electronic spectra in the Sm side which presented an additional energy gap to the tunneling of electrons. Very recently, a theoretical model also studied the effect of the Schottky barrier present at the Sm-Sc interface.[31] Their results predict asymmetry in the conductance spectra and huge enhancements in the Andreev signal due to resonant tunneling for an appropriate barrier.

Observation of phase coherent phenomena in Sm-Sc junctions through transport measurements was restricted to complex geometry with elaborate fabrication techniques. In this paper, we report soft point contact Andreev reflection spectroscopic studies between highly doped n type Si (n ++ Si) and In[32]. PCAR spectra or conductance vs bias voltage *(G(V) = dI/dV vs V)* spectra of the n++Si-Indium (In) junction shows clear yet broad spectral gap features at low temperatures (Here, the spectra resembled those in the tunneling regime rather than the point contact regime as discussed in Ref[29]). For most of the contacts, these broad spectral features with distinct coherence peaks were also observed at temperatures which were as high as about 80% of the transition temperature ($T_c$) of In. Analyzing the spectra measured at the lowest temperatures (~ 1.7 K) using the BTK model, a high value of the superconducting energy gap ($\Delta$) for In was obtained (~1.33 times the BCS gap of In). Also the parameter characterizing the transparency of the contact ($z$ ~ i.e. proportional to the interface barrier height, $V_0$) remained high ($z > 1$) indicating that the transparency was low. However, analysis of the temperature dependence of the PCAR spectra gave the values of $\Delta(T)$ which seemed to follow a BCS variation, with the gap closing at the $T_c$ of In. Interestingly, such anomalous features were *not* observed in soft point contact spectra between Al-In junctions which mimicked a normal metal-superconductor (N-S) junction (Note: Al film was measured above 1.2 K, $T_c$ of Al). This indicated that the observations for the n++Si-In junction were reflective of a Sm-Sc interface. Furthermore, tip induced superconductivity reported from point contact measurements in highly doped Si with Ag tip[33] was also not observed in the present study. Interestingly, such broad spectral features were observed from hard point contact measurements between n++Si and a Pb tip, the contacts again being of low transparency. However, for the contacts studied for the n++Si-Nb junctions, the PCAR spectra were more in the point contact regime indicating that the transparency of the contact was high. Here, the spectra were similar with previous reports which was consistent with the theory of proximity induced interface superconductivity (PIIS) in a Sm-Sc junction, showing a smaller gap in addition to the gap feature associated with the superconducting Nb gap (originating from Andreev reflection)[9]. We analyzed our data based on the model presented in Ref. [30] and [9]. The broad spectral features of the n++Si-In contacts and the n++Si-Pb contacts could be explained on the basis of the presence of the Schottky barrier present at Sm-Sc interfaces consistent with the model of Ref [30]. Our experimental findings indicate that the transparency of the contact is vital in understanding the transport across Sm-Sc interfaces.

**Experimental details**

Commercially available silicon crystals (n type) were used for the measurements presented here. The type and level of doping were confirmed by *in-house* Hall measurements at different temperatures between 100 – 2 K. These doped crystals had a resistivity of ~0.9 mΩ cm, with a doping concentration ~3-5x10$^{25}$ per m$^3$ as seen from the Hall data at 25 K in Fig. 1(b). It is worth mentioning that the mean free path in the n++ Si is ~ 12 nm and the coherence length, $\xi_{Sm}$ is ~ 19 nm at 1.7 K ($\xi_{Sm} = \sqrt{\frac{\hbar v_F l}{6\pi T k_B}}$, where $v_F$ is the Fermi velocity)[9]. Soft point contacts (PC) were made with a small piece of indium pressed on the Si crystals. The piece of Si was mounted on a glass slide and contacts to the In (superconducting with $T_c$ = 3.4 K, see inset of Fig. 3(a)) was made by fine Au wires. Cu wires were used to make contacts on the Si (schematic shown in Fig. 1(a)). Transport measurements of the n++ Si/In junction was done by the standard four probe configuration. A Keithley current source was used to bias the junction and a nanovoltmeter was used to measure the voltage drop. The *i-v* curves were numerically differentiated to get the conductance spectra (dI/dV vs V). In soft PC though the macroscopic contact area is large, the effective electrical contact happens over a much smaller area due to the presence of parallel micro bridges in the contact area. Contacts in both the ballistic (where the effective contact diameter, *d* << elastic mean free path of the electrons, $l_{el}$) and diffusive regime (where *d* << ($l_{el}$ $l_{in}$)/3)$^{0.5}$, $l_{in}$ being the inelastic mean free path) can provide useful and same energy resolved information from the PC[34,35]. Microscopic area of the contact was tuned by applying small voltage pulses. The contact resistance of the soft PC data presented here was about 2 – 20 ohm which corresponded to the apparent contact diameter of about tens of nanometers[34]. The effective contact diameter is therefore expected to be smaller implying that the transport is likely to be in any one of the two regimes, ballistic or diffusive. Since the actual microscopic PC is unknown, standard diagnostics based on the shape of the conductance spectra were done to ensure that the spectra were not in the thermal regime[32,36,37]. In the thermal regime, conductance dips are observed at biases, $V \gg \Delta/e$ where $\Delta$ is the superconducting energy gap. These spectral features are associated to the local heating at the contact which lead to the bias current reaching the critical current of the superconductor[37]. By slowing tuning the contact, and increasing the contact resistance, the contact diameter can be decreased to enter the non-thermal regime of transport where these dips disappear and the characteristic double peak symmetric about $V = 0$ distinctly appear. Since, the $l_{el}$ is quite low in n++ Si, it is more likely that the transport is in the diffusive regime as was reported in similar measurements on n++Si-Nb junctions[9]. Hard point contact on Si was done by the usual needle-anvil technique with different tips like Nb, Pb, Ag, Cu, Pt-Ir etc. Pressure on the contacts was adjusted at room temperature to keep the normal state resistance to a few ohms. As explained above, similar diagnostics were done to eliminate any spectra in the non-thermal regime.

**Results and Discussions**

Figs 1(c) shows the PCAR spectra acquired by soft PC on a highly doped n-type Si at a temperature $T < T_c$ of In. Three distinct observations emerge from the data: i) Andreev reflection is occurring at the Sm-Sc interface leading to the characteristic double-peak structure of a PCAR spectra (looking at the spectra the contacts look more in the tunneling regime)[29] ii) Unlike the symmetric double peak structure seen in a typical PCAR spectra for N-Sc interfaces, the spectra in Fig. 1 (c) are asymmetrical. Notably the asymmetry decreases on decreasing the normal state contact resistance, $R_N$ or increasing conductance *(dI/dV)* (see lower panel in Fig. 1(c)) iii) the coherence peaks are quite pronounced even at temperatures as high as ~ *0.8T_c*, where $T_c$ is the transition temperature of the superconducting In. and iv) the spectra are relatively broad with peak to peak voltage ~ 2.0-2.3 mV.

The asymmetrical nature of the PCAR spectra can be understood on the basis of the recent work by Bouscher *et. al.* who developed a theoretical model to understand the conductance of Sm-Sc junctions with arbitrary potential barriers[31]. Their results show that due to the presence of the Schottky barrier between the degenerate Sm and the Sc, the process of the retro-reflection of the quasi-particles become non-ideal and hence leads to an asymmetry in the conductance spectra. This asymmetry was also seen in earlier experiments of Si-Nb contacts where the pair current was observed from the transport measurements of the junctions[3]. In our data the decrease in asymmetry of the PCAR spectra with decreasing contact resistance can be understood from the fact that with tuning the contact with local voltage pulses, the potential barrier at the junction reduces. In Fig. 1(d), PCAR spectra are shown for different contact resistances, $R_N$. In these set of spectra, the asymmetry is removed manually by normalizing each spectrum with the spectra obtained at $T > T_c$ for each contact. From these spectra, in addition to the above mentioned four distinguishing features, we can see that with decreasing $R_N$, the coherence peaks diminishes considerably. Following the work of Heslinga *et al.*[9], we fit the spectra with the BTK model to obtain the values of the superconducting energy gap ($\Delta$) of bulk In. A broadening parameter $\Gamma$ was introduced in the BTK model to account for all sources of non-thermal broadening[38]. Besides, $\Delta$ and $\Gamma$, the interface transparency was modelled with a dimensional parameter $z$ which is taken to be proportional to the potential barrier potential, $V_o$ at the N-Sc interface in the BTK model. For most of the spectra, $z$ ranged between 1.0-1.7, indicating that the contact was primarily in the tunneling regime. This can be explained on the basis of the large mismatch between the Fermi levels between the Sm-Sc which reduces the transparency at the interface. The values of the $\Delta$ and $\Gamma$ obtained from fitting the PCAR spectra with different $R_N$ is shown in Fig. 1(e) and also in Table I. [Note: Some of the spectra analyzed whose data have been presented were measured at a base temperature of *T = 1.7 K* and many of the spectra were measured at a base temperature of *T = 2.8 K*]. From Fig. 1(e), we can see that for some of the contacts with $R_N$ < 6 ohm, $\Delta$ becomes almost comparable to $\Gamma$, which explains the observed broadened PCAR spectra for these contacts as shown in Fig. 1(d). Furthermore, it appears that the value of $\Delta$ for In is about 0.8 meV [obtained at $T = 0.52\ T_c$]. Indium being a BCS superconductor, the expected value of $\Delta(0) = 0.6\ meV$ [$\Delta(0) \sim 2\ k_BT_c$] and hence the experimental value of the superconducting energy gap appears to be quite large (about 1.3 times the bulk In gap). For low resistance contacts which appeared more broadened, a careful look at the spectra shows additional weak features appearing symmetrically about the zero bias indicating that those could be real and not mere artifacts. Figure 1(f) shows the spectra with $R_N$ = 5.9 ohm. Here, the first feature appears at a voltage, $V_1$ ~ 1.3 mV which is still quite high if it is to be associated with the gap of In. The additional second feature is seen at a bias voltage of $V \sim 2V_1$. To understand these anomalous features in the PCAR spectra, it becomes pertinent to ask, what is the origin of the large gap of In observed for all contacts with highly doped Si. Can these observations be explained i) simply as experimental artefacts in soft PC spectroscopy or ii) by the fact that highly doped Si might show tip induced superconductivity (TIS) consistent with the recent report by Sirohi *et al*[33] or iii) on the basis of the Schottky barrier present at the Sm-Sc interface[30] or iv) by the presence of a second superconducting phase appearing at the Sm-Sc interface induced by the proximity effect[9].

Possibility (i) of experimental artefacts in soft PC spectroscopy can happen since a relatively large area contact is made by pressing a piece of In on the metal or semiconductor. It might be doubtful whether the PCAR spectra can be modelled by the BTK theory and hence the large values of the gap obtained from fitting of the experimental spectra can arise. However, it should be reiterated that this technique has been successfully demonstrated to give correct values of the superconducting energy gap for films of Nb, Pb and NbN *etc*[32]. It is well established that in soft PC technique (either with In or Ag), several point contacts are formed at the interface and the transport can be successfully described by the BTK model[39]. In order, to re-emphasize that this is not due to any experimental artefact, we carried out soft PC spectroscopy on Al film with In at temperatures greater than the $T_c$ of Al (~ 1.2 K). This behaved like a normal N-S contact. The PCAR spectra (shown in the upper panel of Fig. 2(a) was fitted with the BTK model. The values of $\Delta(T)$ followed the BCS variation as seen from the lower panel of Fig. 2(a). Besides, $\Delta(0)$ was consistent with the BCS value of gap for bulk indium. This proves that the large gap obtained for In is either the result of the Sm-Sc contact or is particular to the interface of the highly doped Si.

The basis of the possibility (ii) of TIS is based on the recent report by Sirohi *et. al.* on PCAR and scanning tunneling spectroscopy (STS) studies on highly doped Si probed with non-superconducting tips where clear signatures of superconductivity with a $T_c$ of 10 K was observed[33]. The authors attributed it to a tip-induced superconducting (TIS) phase in doped silicon which has been observed more frequently in systems with topologically non-trivial band structures. To check, if our results can be explained on the same basis, we carried out PC measurements by the conventional needle-anvil technique or hard point contact on the n++ Si using normal metal tips. We present the data with Ag tip in Fig. 2(b) where we obtained a negative result. From both the temperature variation of $R_N$ of the contact (upper panel) as well as the conductance spectra (lower panel), no signatures of superconductivity in n++Si was observed. This was true for most of the other normal tips we probed like Pt-Ir and Cu. Hence, we rule out the possibility of TIS in the n++ Si interface in our experiments.

Next we explore the possibilities (iii) and (iv). We measured the temperature dependence of the normal state resistance, $R_N$ of the soft PC which is shown in Fig. 3(a) for a contact with $R_N = 8.0$ ohm. A single drop in the PC resistance is seen at $T = 3.4$ K corresponding to the $T_c$ of In (see inset of Fig. 3(a)). The temperature variation of the PCAR spectra for this contact is shown in Fig. 3(b). The PCAR spectra seemed to become completely featureless at $T > 3.4$ K indicating a closure of the gap at the $T_c$. Each spectrum was analyzed using the modified BTK model to obtain $\Delta(T)$ which is plotted as a function of temperature in Fig. 3(c) as red solid circles. We have similarly analyzed spectra for another contact with $R_N = 18$ ohm. The obtained $\Delta(T)$ from the fits have been shown in Fig. 3(c) as blue squares. Also, shown in this plot is the BCS variation of the gap for In (green dashed line). As is clearly seen that the values of $\Delta(T)$ are consistently higher at all temperatures than that expected for Indium. However, interestingly, the temperature variation of the gap values of In obtained from the fits mimics the BCS variation (as seen by the black dashed lines in the figure), albeit with extremely high ratio of $\Delta/T_c$.

We next explore possibility (iii) in greater detail on the basis of the work reported in Ref [30]. It has been shown there that the presence of a Schottky barrier at the Sm-Sc interface results in a reduction of the charge screening along the Sm side. This leads to an additional energy gap, $E_B$ which along with the superconducting energy gap, $\Delta$ presents a larger energy gap of $(E_B + \Delta)$ to single electron tunneling. Consequently, the conductance spectra of the interface shows peaks at voltages of $(E_B + \Delta)/e$ at temperatures $T \ll T_c$. We analyze our data based on this model. According to the model, assuming that in the Sm side, states get depleted in an energy range of $E_B$ across the Fermi level (μ), $E_B$ can be evaluated using the equation:

$$\int_{\mu-E_B/2}^{\mu+E_B/2}(\varepsilon - E_C)^{1/2} d\varepsilon = nk_F^{-1}/2\pi L^* \left(\frac{2m_n^*}{h^2}\right)^{3/2} \tag{1}$$

Here, $L^*$ is the phase coherence length across the interface and is usually the minimum of the inelastic scattering length, $l$ and the thermal length, $l_T$ ($l_T = (D\hbar/2\pi k_B T)^{1/2}$, where $D$ is the Diffusion constant). $n$ is the carrier concentration, $k_F$ is the inverse of the screening length which can be evaluated for any Sm-metal interface from the depletion layer width, $w$, $m_n^*$ is the effective mass of the electrons and $E_C$ is the conduction band edge. Now, for the degenerate Sm with a carrier concentration of $n$, $(\mu - E_C)$ can be determined from:

$$\mu - E_C = \left[\left(\frac{3}{8\pi}\right)n\right]^{2/3} \frac{h^2}{2m_n^*} \tag{2}$$

Taking, $n = 3.8 \times 10^{25}$ /m$^3$ as determined from the Hall measurements for the n++Si and $m_n^*$ as $0.98m_0$ where $m_0$ is the rest mass of electrons, $(\mu - E_C)$ comes out to be 43 meV. $(\mu - E_C)$ can be incorporated in equation 1 to give:

$$\left[2\pi\left(\frac{2m_n^*}{h^2}\right)^{\frac{3}{2}}/n\right]\int_{-E_B/2}^{E_B/2}(\xi + (\mu - E_C))^{1/2}\,d\xi = (L^*k_F)^{-1} \tag{3}$$

Using the above equation, we plot the variation of $E_B$ with $(L^*k_F)^{-1}$ in Fig. 3(d) for $(\mu - E_C)$ of 43 meV. Furthermore, the depletion layer width can be estimated from, $w = \sqrt{\frac{2\varepsilon_0\varepsilon_r(\Phi_B - V_{th})}{qn}}$ where, $\varepsilon_0$ is the permittivity of free space, $\varepsilon_r$ is the dielectric constant of the semiconductor (~11.9 for Si), $\phi_B$ is the barrier potential and $V_{th} = 2k_BT/q$. Now, for the n++Si-In junction, $(L^*k_F)^{-1}$ is about 0.01 which gives an $E_B$ of about 0.58 meV (see fig. 3(d)). Moreover, a rough estimate of $E_B$[30], given by $E_B = \frac{\hbar^2}{2m_n^*w^2}$ gives a value of 0.9 meV. With $\Delta(0)$ for In ~ 0.6 meV, at $T<<T_c$, this would lead to peaks in the conductance spectra at 0.8 to 1 mV which is consistent with the observed values shown in Figure 3 (b). Furthermore, this model can also explain quantitatively the position of the peak voltage shown in Figure 1(d) for $T = 0.8T_c$ by accounting for the temperature variation of the In gap and the Fermi distribution. The variation in the gap values observed with contact resistance (Fig. 1(e)) can also be understood on the basis of the variation of the interface Schottky barrier with contacts which would affect $E_B$. The different parameters for different contacts (different $R_N$) obtained from the analysis is shown in Table 1. We estimated $E_B$ by three different ways. $E_B^*$ was obtained from (eV$_{pk}$ - $\Delta_{Sc}$) where $V_{pk}$ was the peak position of the coherence peak observed in the PCAR data and $\Delta_{Sc}$ was the expected superconducting energy gap of In at the measurement temperature, T. $E_B^{\#}$ was simply obtained from ($\Delta_{BTK} - \Delta_{Sc}$), where $\Delta_{BTK}$ was the value of the gap obtained by fitting the experimental PCAR spectra by the BTK model. Finally, $E_B$ was obtained from figure 3(d) and using the crude expression of $E_B$ mentioned above. Interestingly, a fairly good match was obtained between $E_B*$ and $E_B$. Thus, the model of the charge screening by the Schottky barrier not only explains our data for the n++Si-In contacts qualitatively, but also reasonably good quantitative match is obtained.

Next, we investigated the magnetic field evolution of the PCAR spectra at different temperatures for the n++Si-In contacts. PCAR spectra was measured at a constant temperature for different magnetic fields at temperatures of T = 1.9 K, 2.1 K, 2.4 K, 2.6 K and 2.8 K. The spectra obtained at T = 1.9 K is shown in Fig. 4(a). Each spectrum was fitted with the single gap BTK model to extract $\Delta_s(T)$ for different magnetic fields. This is plotted for different temperatures in Fig. 4(b). At the lowest measured temperature of 1.9 K, the critical field where the gap would close seems to be greater than 3.5 kG. We attribute it to In though it seems to be substantially enhanced from 0.28 kG for bulk In[40]. This is not surprising as critical fields are known to be enhanced at meso-scopic junctions[41]. Furthermore, the magnetic field variation for higher temperatures seems more or less linear.

To further check the effect of the Schottky barrier on the peak position of the PCAR spectra, we explored interfaces of n++Si with other superconductors like Pb and Nb. We carried out hard point contact spectroscopy with these superconducting tips on n++Si. These experiments would also help us investigate the final possibility i.e. (PIIS) and how it influences PCAR spectra. Since the $T_c$ of both these superconductors are higher than In, any signature of another superconducting phase (even weak) at the interface would result in well resolved peaks in the PCAR spectra. It is worth mentioning that Andreev reflection studies done on n++Si - Nb junctions[9] as well as hetero-structures of many topological materials like $Bi_2Se_3$, $Bi_2Te_3$ *etc*[12,13] with superconductors have shown the presence of a sub gap feature which is related to the proximity induced gap at the interface of the Sm/topological semi-metals. These sub-gap features are in addition to the gap features arising from Andreev reflection. Fig. 5(a) shows the temperature variation of the PCAR spectra for a n++Si - Pb junction. The spectra became featureless at the *$T_c$ = 7.0 K* of the point-contact (As seen from the temperature variation of $R_N$ of the contact in the inset). However, similar to the n++Si-In junctions, the contact appears to be in the tunneling regime and gives broad spectral features. The peak to peak voltage was 5.0 mV. This data can also be analyzed based on the model of the Schottky barrier. For the n++Si-Pb interface, $E_B$ is expected to be ~ 0.78 – 1.28 meV which can again explain quantitatively the observed peak position in the PCAR spectra (See Table 1). For another contact shown in Fig. 5(b), (though $R_N$ was not very different), which seemed more in the PC regime (low *z*), the broad spectral feature seemed to show multiple very weak humps. The first hump was observed at $V_1$ = 1.3 mV which matched closely with the expected gap feature for Pb. The second peak which was the most prominent peak was observed at $V_2$ = 2.5 mV. Besides, a small peak was also seen at zero bias. If we neglect the extremely weak hump at $V_1$, the peak position at $V_2$ can also be explained on the basis of the Schottky barrier which would give rise to an $E_B$ ~ 1.3 meV.

Results of similar studies with Nb tip is shown in Figs. 5(c) – (d) for two different contacts. Both contacts seem to be in the PC regime with high transparency. Two striking difference can be observed in these spectra when compared with those for the Si-In and Si-Pb contacts. First, the coherence peaks are seen at the bias voltage corresponding to the gap feature of Nb i.e. at V = 1.5 mV. The spectra in Fig. 5(c) was analyzed on the basis of the BTK theory as seen from black solid line in Fig 5 (c). The other prominent difference were the presence of sub gap or mini gap feature seen at V = 0.5 – 0.7 mV (depending on $R_N$). To understand the spectra for this high transparency (low *z*) contact, we invoked the theory of multiple Andreev reflection (MAR) prevalent for a Sc-Sc contact[42]. Under MAR, PCAR spectra shows peaks at $\Delta_1$, $\Delta_2$, $\Delta_1 + \Delta_2$ etc[32]. The sub gap features in Fig. 5(c) and (d) might correspond to another superconducting phase at the interface with a gap of $\Delta_1$. The origin of the second superconducting phase could be due to proximity induced interface superconductivity (PIIS) of the n++Si - Nb junction similar to that seen in many Sm-Sc junctions[9]. It is worth mentioning PIIS give rise to the zero bias peak (seen in Fig. 5(c)) associated with Andreev scattering known to occur in very transparent Sm-Sc junctions. Thus, the presence of the prominent peak at a voltage corresponding to the gap of the Sc and the sub gap features clearly indicates that these low $z$ contacts cannot be analyzed on the basis of the Schottky gap (see Table I). Also, from Fig. 5(d), both gap features evolve with temperature with the larger feature closing at the $T_c$ of the Sc. This further indicates that PIIS at the Sm-Sc interface is responsible for the observed features in the PCAR spectra for high transparency junctions of n++Si-Nb junctions.

**Conclusions**

In conclusion we report on the point contact Andreev reflection (PCAR) studies on highly doped n type Si. Both soft PCAR with In and hard PCAR spectroscopy with superconducting tips like Nb and Pb was done on Si. The soft PCAR spectra on Si seemed considerably broadened with enhancements in the gap value of In from its BCS bulk value, though it closed at the $T_c$ of In. Similar result was also obtained from some of the contacts in hard point contact studies with Pb tips which were in the tunneling regime. However, PCAR spectra of n++Si - Nb junctions which were more transparent showed distinct signatures for PIIS, which resulted in a sub-gap feature in addition to the distinct gap feature from Sc Nb. The low transparency contacts (for n++Si-In and n++Si-Pb) can be analyzed on the basis of the Schottky barrier present at the Sm-Sc interface. In this model, an additional energy gap, $E_B$ arises in the Sm side due to in-efficient charge screening which increases the bias at which the coherence peaks are obtained in the conductance spectra to $\Delta + E_B$ where, $\Delta$ is the superconducting energy gap. A quantitative treatment for the n++Si-In junction gave an $E_B$ of 0.58 to 0.9 meV which explained the experimental PCAR spectra quite well. Our experimental findings convincingly show that based on the transparency of the contacts for Sm-Sc junctions, the PCAR spectra is seen to be influenced differently. For low transparency contacts, the Schottky barrier primarily influences the transport resulting in shifting of the coherence peaks from $\Delta/e$ to higher biases while for highly transparent junctions, proximity induced interface superconductivity dominates the transport giving distinctive sub gap features and zero bias peak in the conductance spectra.

**Acknowledgements**

We will like to thank Mr. Soumyajit Mandal and Prof. P. Raychaudhuri for the Hall measurements. SB acknowledges partial financial support from the Department of Science and Technology, India through No. SERB/F/1877/2012.

**Author Contributions**

PP carried on the experiments and analyzed the data. SB conceptualized and supervised the project and also refined the analysis by invoking the model of a degenerate semiconductor and a superconductor. Both authors discussed the results and SB wrote the manuscript.

**Table I: Comparison of $E_B$ from experimental data and theoretical model for different contacts and for different junctions of n++Si with different superconductors (Sc).**

| Sc | Regime | T (K) | $R_N$ | Δ (BTK fit) (meV) | Γ (BTK fit) (meV) | z (BTK fit) | $\Delta_{Sc}$ BCS (meV) | $V_{pk}$-$V_{pk}$ (From Expt.) (mV) | $E_B$* (From Expt. data) (meV) | $E_B$# (From BTK fit) | $E_B$ (From model) |
|---|---|---|---|---|---|---|---|---|---|---|---|
|  | Tunneling | 2.85 | 2.4 | 0.65 | 0.65 | 1.1 | 0.4 | 2.56 | 0.48 | 0.25 | ~0.5-0.9 meV |
|  | Tunneling | 2.85 | 3.0 | 0.70 | 0.62 | 1.65 | 0.4 | 2.60 | 0.52 | 0.30 |  |
| In | Tunneling | 2.85 | 3.3 | 0.68 | 0.52 | 1.5 | 0.4 | 2.6 | 0.60 | 0.28 |  |
|  | Tunneling | 2.85 | 4.7 | 0.68 | 0.62 | 1.5 | 0.4 | 2.80 | 0.62 | 0.28 |  |
|  | Tunneling | 2.85 | 5.9 | 0.70 | 0.54 | 1.2 | 0.4 | 2.54 | 0.56 | 0.30 |  |
|  | Tunneling | 2.80 | 10.4 | 0.52 | 0.48 | 1.7 | 0.41 | 2.54 | 0.59 | 0.11 |  |
|  | Tunneling | 1.78 | 8.0 | 0.80 | 0.45 | 1.1 | 0.59 | 2.2 | 0.39 | 0.21 |  |
|  | Tunneling | 2.40 | 18.0 | 0.80 | 0.35 | 1.0 | 0.51 | 2.3 | 0.45 | 0.29 |  |
| Pb | Tunneling | 2.8 | 3.9 | BTK fit not possible | | | 1.5 | 5.0 | 1.0 | - | ~0.78-1.3 meV |
|  | Point contact | 2.4 | 5.0 | BTK fit not possible | | | 1.5 | 2.6 | - | - |  |
| Nb | Point contact | 2.6 | 7.9 | 1.42 | 0.65 | 0.45 | 1.5 | 3.0 | - | - | ~0.84-1.43 meV |
|  | Point contact | 2.8 | 7.5 | BTK fit not possible | | | 1.5 | 3.0 | - | - |  |

**Figure Captions:**

**Figure 1:**

(a) Schematic of the soft point contact with In pressed on highly doped n type Si. (b) Hall resistance *vs* magnetic field at T = 25 K for the n++ Si showing a doping of ~ 3.85 x $10^{25}$ /m$^3$. (c) Soft PCAR spectra for n++ Si for different contact resistances at T < 3.0 K. (d) Soft PCAR spectra (Solid circles) for n++ Si for the contact resistances $R_N$ = 3.0, 5.9, 8.0 and 18.0 ohms. The solid lines show the fit according to the BTK theory. (e) Variation of Δ (red and magenta solid circles) and Γ (blue solid squares) with contact resistance obtained from fitting the PCAR spectra using BTK theory. (f) PCAR spectra for $R_N$ = 5.9 ohm show peaks at $V_1$ and a mild hump at $V_2$ (~$2V_1$).

.

**Figure 2:**

(a) The top panel shows the soft PCAR spectra on Al film with pressed In. Solid symbols are data and the solid lines are fits with the single gap BTK model. The lower panel shows the temperature variation of the gap obtained from the fits which fits well with the BCS variation for In with $T_c$ = 3.45 K. (b) Top panel shows the temperature variation of the contact resistance, $R_N$ for a hard point contact between n++ Si and Ag tip. The lower panel shows the differential conductance of the junction indicating absence of any signatures of superconductivity of the n++ Si.

**Figure 3:**

(a) Temperature dependence of the contact resistance, $R_N$ of the soft PCAR spectra with n++ Si showing a single drop at T = 3.4 K. Inset shows the temperature variation of real part of the susceptibility for a thick In film (~ 200 nm) grown using the same In used for PCAR measurements showing the critical temperature ($T_c$) as 3.4 K. (b) Normalized Soft PCAR spectra (normalized to the normal state value of the conductance) for n++ Si for the same contact as seen in (a) with $R_N$ = 8.0 ohms for different temperatures from 1.78 K to 3.2 K. A broad coherence peak is obtained. The solid circles are the experimental data while the solid lines are the fits using the single gap BTK theory. (c) Temperature dependence of the superconducting energy gap, Δ(T) (red solid circles) obtained from the BTK fits shown in (b) for the contact with $R_N$ = 8.0 ohms. The blue solid squares are the Δ(T) obtained from analyzing PCAR spectra for another contact with $R_N$ = 18.0 ohms. The black dashed line is the expected

BCS variation of the superconducting energy gap of In. Both the experimental gap variation is higher than the expected In gap for all temperatures. (d) Plot of $E_B$ vs $(k_F L^*)^{-1}$ for the n++Si-In contacts where $(E_c-\mu)$ is 43 meV for the doping of $3.8 \times 10^{-25}$ /m$^3$ for the n++ Si.

**Figure 4**

(a) "Soft" PCAR spectra at different magnetic fields at T = 1.9 K for n++ Si. The open circles are data and the solid lines are fits with the single gap BTK model. (b) Magnetic field variation of gap ($\Delta$) at different temperatures below the $T_c$ of In. $\Delta$ is obtained from analyzing the PCAR spectra using the BTK model measured at a constant temperature for different magnetic fields.

**Figure 5**

(a) PCAR spectra at different temperatures for n++Si – Pb contact with $R_N$ = 3.9 ohm. The two vertical lines indicate the peak to peak voltage of 5.0 mV. Inset shows the temperature variation of $R_N$ for the same contact showing an abrupt increase at a T = 7.2 K. (b) PCAR spectrum for another contact on same n++Si with Pb tip at T = 2.4 K. The spectra show multiple gap features with peak to peak voltages of 2.6 and 5.0 mV. (c) PCAR spectrum (orange solid circles) at T = 2.4 for n++Si-Nb contact with $R_N$ =7.9 ohm showing multiple gap features with peak to peak voltages of 1.4, 3.0 and 4.4 mV. The zero bias conductance peak is indicated by brown dashed arrow. Black solid line is the fitting curve with the help of BTK model which fits the peak at V = 1.5 mV and corresponds to the gap of Nb. (d) PCAR spectra of n++Si-Nb contact with a different contact resistance, $R_N$ = 7.5 ohms at different temperatures. Here, two gaps features are seen which evolve according the dashed arrows (magenta and green) with increasing temperature.

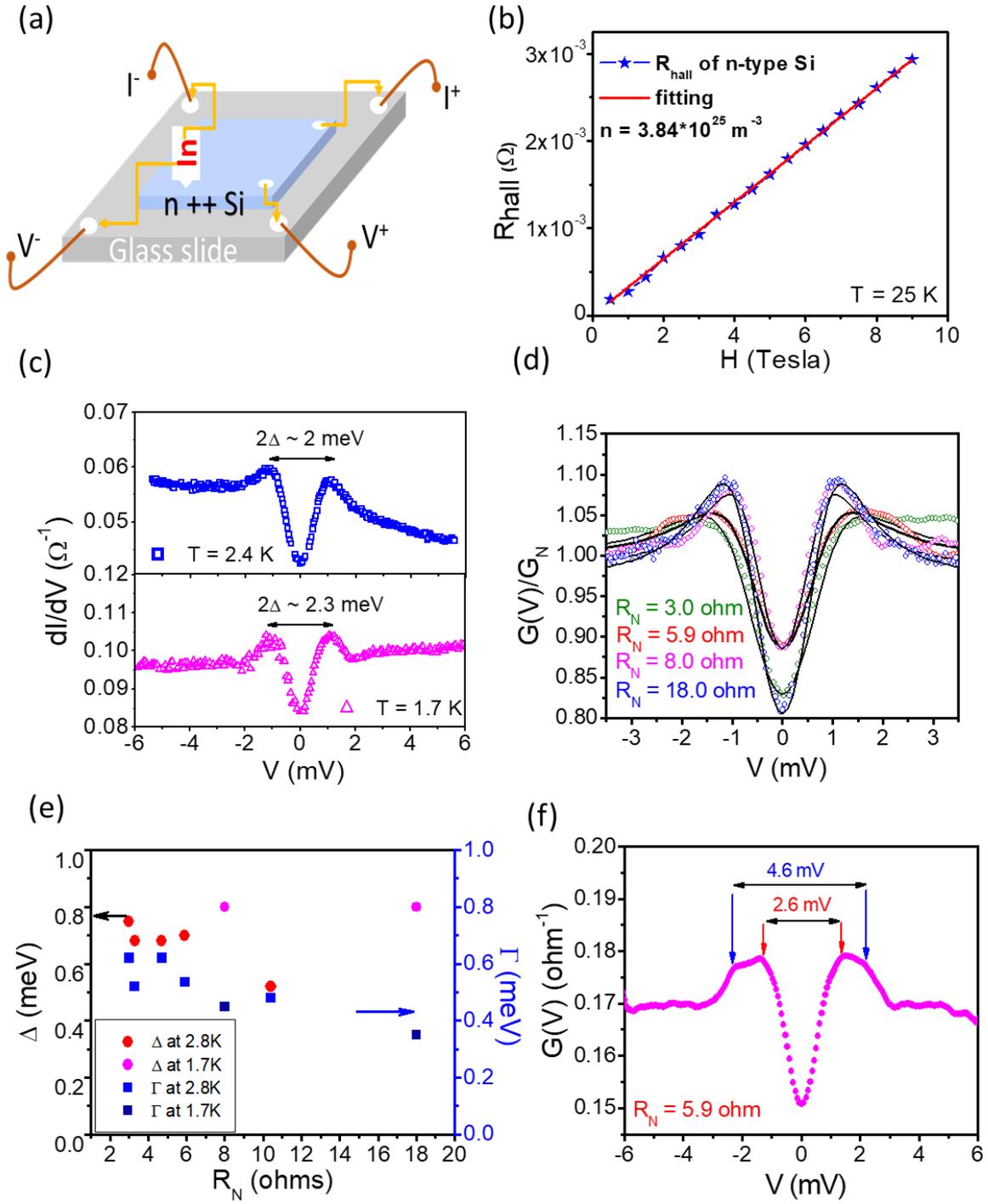

**Figure 1**

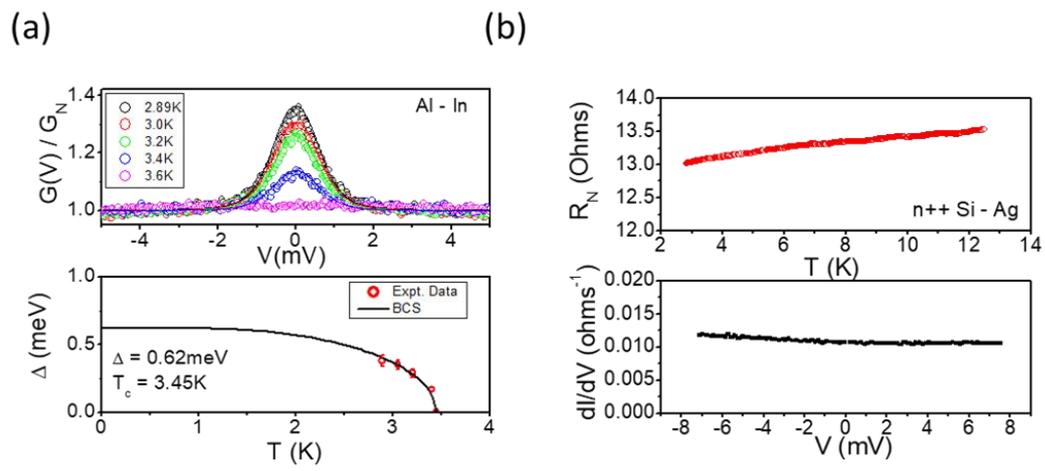

**Figure 2**

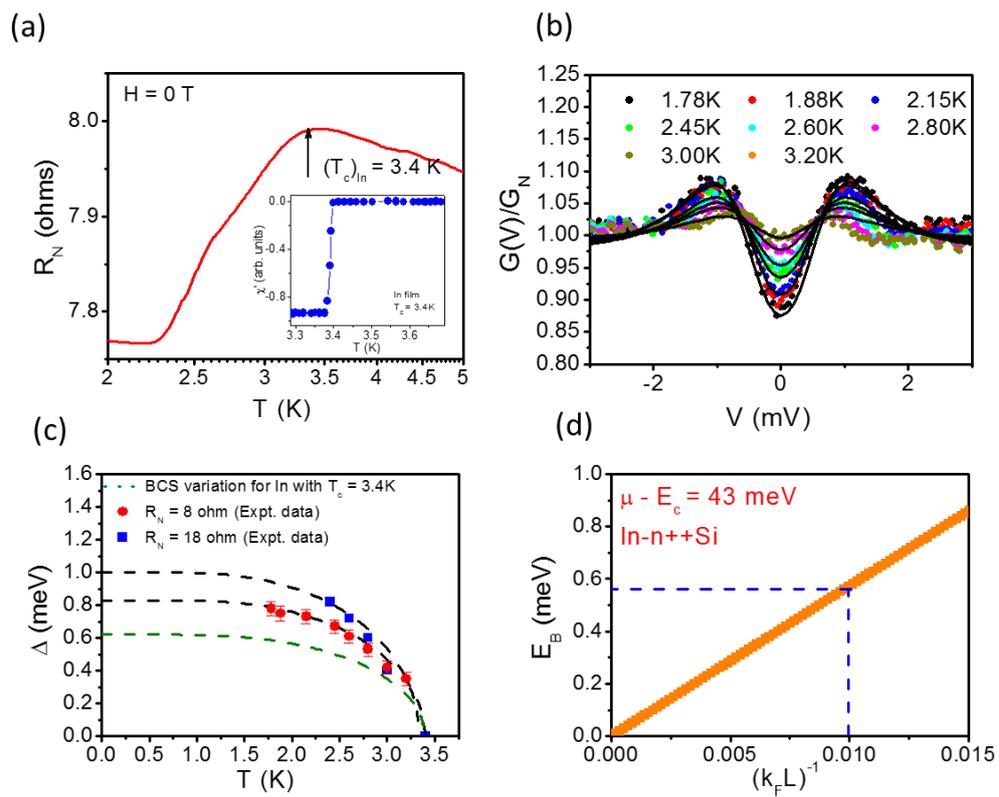

**Figure 3**

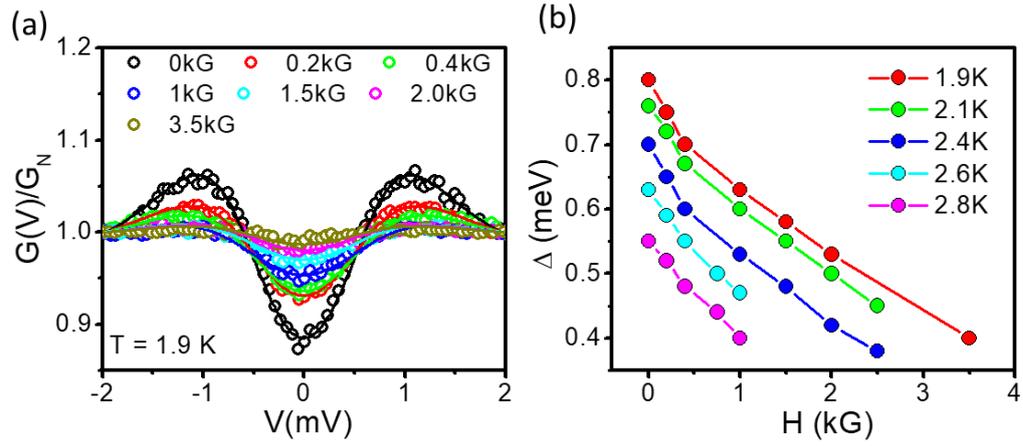

**Figure 4**

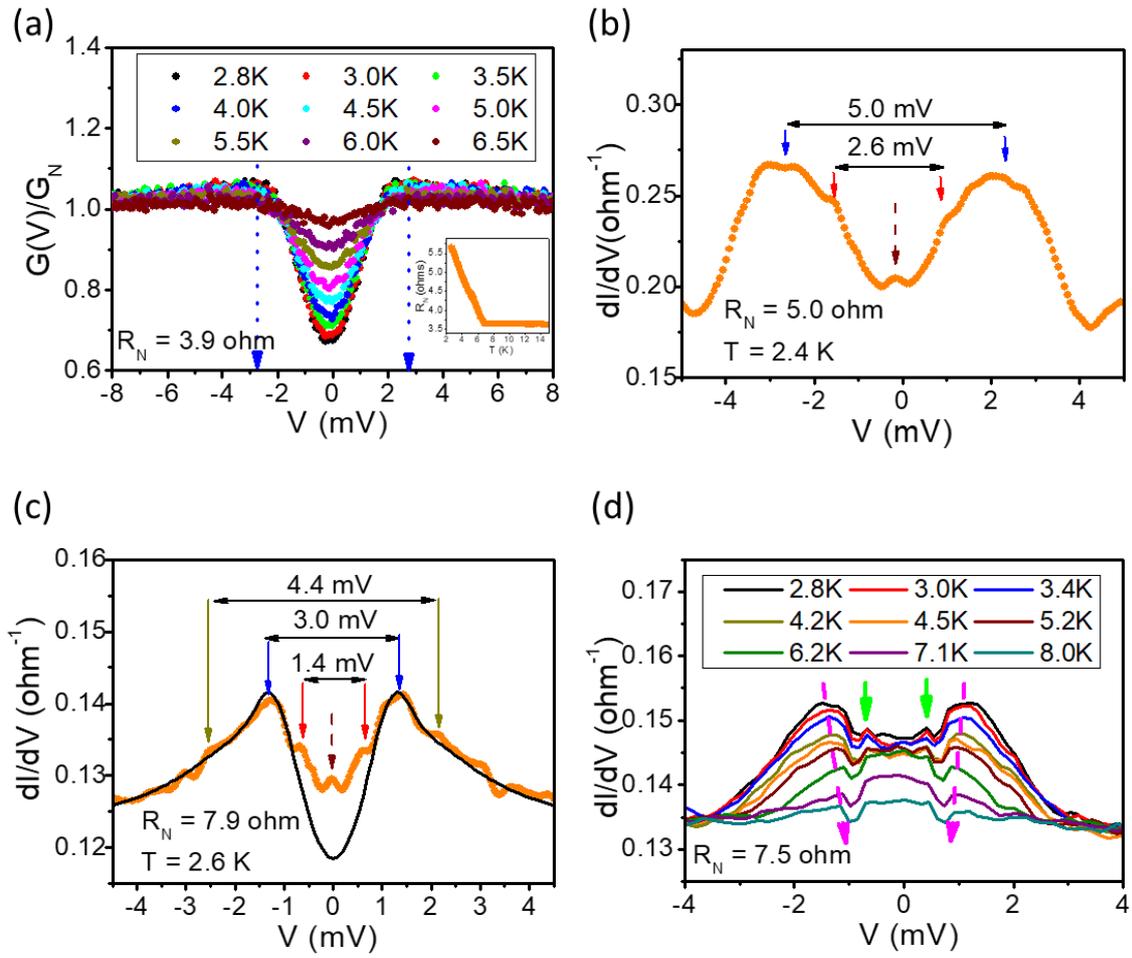

**Figure 5**